\documentclass[sigconf]{acmart}

\AtBeginDocument{%
  \providecommand\BibTeX{{%
    \normalfont B\kern-0.5em{\scshape i\kern-0.25em b}\kern-0.8em\TeX}}}

\usepackage{multirow} 
\usepackage{booktabs}
\usepackage{amssymb}  
\usepackage{tabularx}
\usepackage{makecell}
\usepackage{color}
\renewcommand\footnotetextcopyrightpermission[1]{}
\acmConference[xxx]{xxx}{xxx}{xxx}%

\author{Zengding Liu}
\authornote{Both authors contributed equally to this research.}
\affiliation{%
  \institution{Shenzhen Institute of Advanced Technology, Chinese Academy of Sciences}
  \city{Shenzhen}
  \country{China}}
\affiliation{%
  \institution{University of Chinese Academy of Sciences}
  \city{Beijing}
  \country{China}}
\email{zd.liu@siat.ac.cn}
\author{Chen Chen}
\authornotemark[1]
\affiliation{%
  \institution{Hong Kong Polytechnic University}
  \city{Hongkong}
  \country{China}}
\affiliation{%
  \institution{Shenzhen Institute of Advanced Technology, Chinese Academy of Sciences}
  \city{Shenzhen}
  \country{China}}
\email{christina.chen@connect.polyu.hk}

\author{Jiannong Cao}
\affiliation{%
  \institution{Hong Kong Polytechnic University}
  \city{Hongkong}
  \country{China}}
\email{csjcao@comp.polyu.edu.hk}

\author{Minglei Pan}
\affiliation{%
  \institution{Shenzhen Institute of Advanced Technology, Chinese Academy of Sciences}
  \city{Shenzhen}
  \country{China}}
\email{ml.pan@siat.ac.cn}

\author{Jikui Liu}
\affiliation{%
  \institution{Shenzhen Polytechnic University}
  \city{Shenzhen}
  \country{China}}
\email{liujikui007@gmail.com}

\author{Nan Li}
\affiliation{%
  \institution{Shenzhen Institute of Advanced Technology, Chinese Academy of Sciences}
  \city{Shenzhen}
  \country{China}}
\email{nan.li3@siat.ac.cn}

\author{Fen Miao}
\authornote{Corresponding authors.}
\affiliation{%
  \institution{Shenzhen Institute for Advanced Study, University of Electronic Science and Technology of China}
  \city{Shenzhen}
  \country{China}}
\email{fenmiao@uestc.edu.cn}
\author{Ye Li}
\authornotemark[2]
\affiliation{%
  \institution{Shenzhen Institute of Advanced Technology, Chinese Academy of Sciences}
  \city{Shenzhen}
  \country{China}}
\email{ye.li@siat.ac.cn}

\graphicspath{{./images/}}

\begin{document}

\title{Large Language Models for Cuffless Blood Pressure Measurement From Wearable Biosignals}



\begin{abstract}
Large language models (LLMs) have captured significant interest from both academia and industry due to their impressive performance across various textual tasks. However, the potential of LLMs to analyze physiological time-series data remains an emerging research field. Particularly, there is a notable gap in the utilization of LLMs for analyzing wearable biosignals to achieve cuffless blood pressure (BP) measurement, which is critical for the management of cardiovascular diseases. This paper presents the first work to explore the capacity of LLMs to perform cuffless BP estimation based on wearable biosignals. We extracted physiological features from electrocardiogram (ECG) and photoplethysmogram (PPG) signals and designed context-enhanced prompts by combining these features with BP domain knowledge and user information. Subsequently, we adapted LLMs to BP estimation tasks through instruction tuning. To evaluate the proposed approach, we conducted assessments of ten advanced LLMs using a comprehensive public dataset of wearable biosignals from 1,272 participants. The experimental results demonstrate that the optimally fine-tuned LLM significantly surpasses conventional task-specific baselines, achieving an estimation error of 0.00 $\pm$ 9.25 mmHg for systolic BP and 1.29 $\pm$ 6.37 mmHg for diastolic BP. Notably, the ablation studies highlight the benefits of our context enhancement strategy, leading to an 8.9\% reduction in mean absolute error for systolic BP estimation. This paper pioneers the exploration of LLMs for cuffless BP measurement, providing a potential solution to enhance the accuracy of cuffless BP measurement.
\end{abstract}

\ccsdesc[500]{Computing methodologies~Artificial intelligence}
\ccsdesc[500]{Computing methodologies~Machine learning}
\ccsdesc[500]{Applied computing~Health informatics}

\keywords{large language models, instruction tuning, cuffless blood pressure, wearable biosignals}


\maketitle

\section{Introduction}
\begin{figure}[htbp]
\centerline{\includegraphics[width=0.5\textwidth]{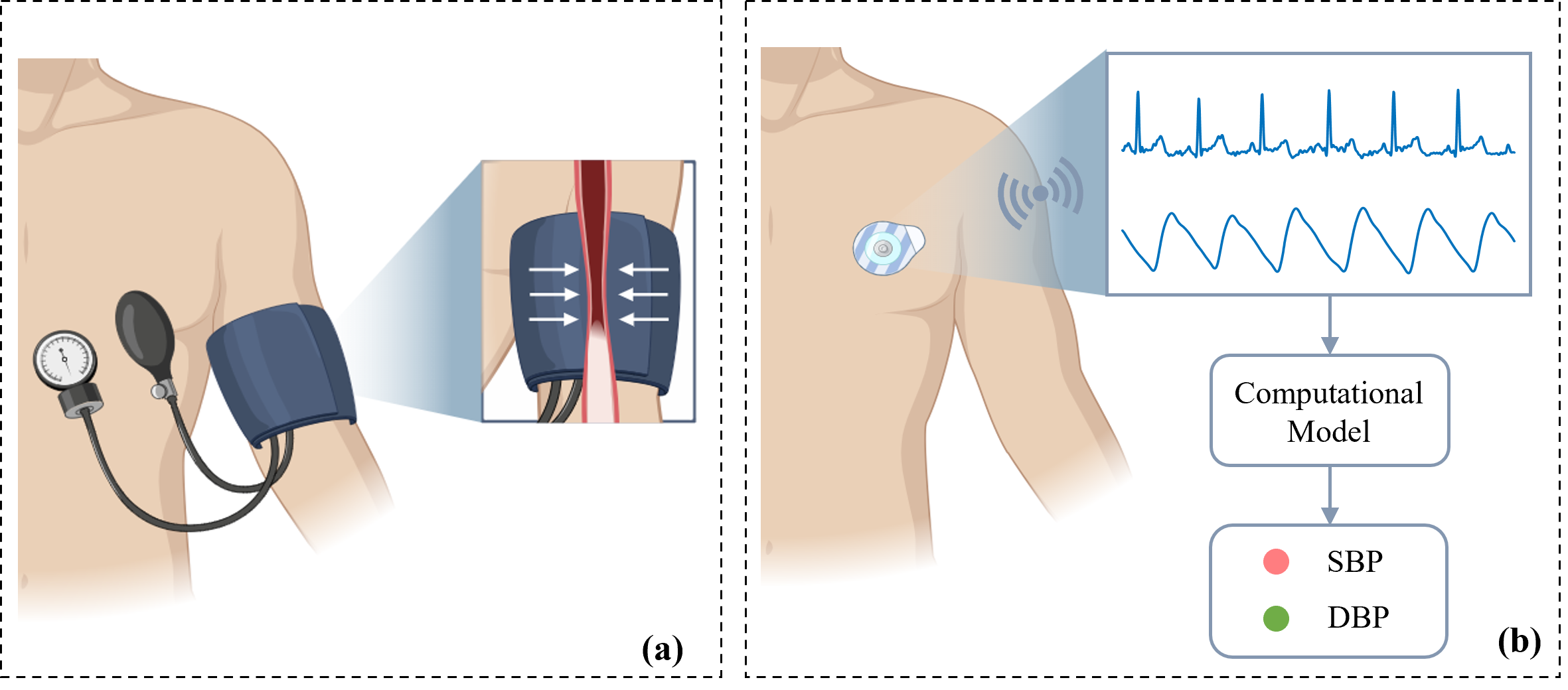}}
\caption{Comparison of cuff-based (a) and cuffless (b) BP measurement approaches. The cuff-based method involves the use of an inflatable cuff to occlude a peripheral artery, which can be uncomfortable. In contrast, the cuffless method calculates BP indirectly from biosignals measured by wearable sensors, avoiding the need for an inflatable cuff. This provides the advantages of comfort and unobtrusive.}
\label{fig1}
\end{figure}

Blood pressure (BP) is an important physiological indicator that reflects the pressure at which the heart pumps blood into the blood vessels. Both high blood pressure (hypertension) and low blood pressure (hypotension) can lead to a variety of health problems, such as heart disease, stroke, and kidney disease \cite{potter2009controlling}. Therefore, accurately monitoring BP is crucial for the prevention and management of various cardiovascular diseases. Traditionally, BP measurement has relied on cuff-based techniques [Figure \ref{fig1} (a)], such as oscillometry and auscultation, which involve the inflation and deflation of a pneumatic cuff around the upper arm. However, these methods suffer from certain limitations, including discomfort, inconvenience, and intermittent measurements.

In recent years, the emergence of cuffless BP measurement techniques has attracted significant interest \cite{ref_1}. These techniques utilize wearable devices to capture biosignals \cite{ref_2}, such as electrocardiogram (ECG) and photoplethysmogram (PPG), and then estimates BP values through advanced computational models [Figure \ref{fig1} (b)]. Compared to the traditional cuff-based method, cuffless BP measurement offers better convenience, unobtrusiveness, and continuous monitoring capabilities, which can bring new possibilities to healthcare and health management. Recent advancements in mechanism-driven pulse transit time models and data-driven machine learning and deep learning models have contributed to progress in cuffless BP measurement \cite{ref_3}. However, further improvement is necessary as these methods have not fully explored the complex patterns of correlation and interactions between biosignals and BP values.

The emergence of large language models (LLMs) provides a new opportunity to enhance the analysis of biosignals. Trained on massive amounts of data, LLMs have demonstrated exceptional capabilities in understanding and capturing complex patterns from data. They have been widely employed in various text-related tasks \cite{ref_4}, such as natural language processing, information retrieval, and content creation, demonstrating strong generalization capabilities. Moreover, LLMs have also been utilized for physiological time series analysis in the healthcare filed, yielding promising results \cite{ref_13, ref_14}. However, the exploration of LLMs' potential in wearable biosignal analysis for cuffless blood pressure measurement remains uninvestigated.

This paper aims to explore the potential of LLMs for cuffless BP measurement using wearable biosignals. The hypothesis is that, through instruction tuning, LLMs can effectively capture the intricate relationships between biosignals and BP values, thereby improving the accuracy of cuffless BP estimation. To achieve this, we extract rich physiological features from the biosignals and embed them as strings into textual prompt templates. By prompt designing and instruction tuning, we construct a model suitable for the cuffless BP measurement task. To the best of our knowledge, this is the first study to utilize LLMs for cuffless BP measurement. Our contributions include:
\begin{itemize}
\item The design of prompt templates using BP domain knowledge to enhance the accuracy of LLMs for cuffless BP measurement.
\item A comprehensive evaluation of 10 state-of-the-art LLMs on a large-scale wearable dataset, demonstrating the potential of LLMs for cuffless BP estimation.
\item The release of CBPM-LLaMA\footnote{Code repository will be made public upon camera-ready.}, an open-source LLM based on LLaMA3-8B specifically fine-tuned for cuffless BP measurement, making it the first of its kind.
\end{itemize}

\section{Related work}

\subsection{Cuffless Blood Pressure Measurement}
Cuffless BP measurement techniques can be broadly categorized into two main types: mechanism-driven and data-driven methods \cite{ref_3}. The most well-known mechanism-driven approach is pulse transit time (PTT), which measures the time for the arterial pulse wave to travel from one arterial site to another by calculating the delay between two biosignals such as ECG and PPG \cite{miao2019multi}. According to the Moens-Korteweg and Hughes equations, an increase in arterial stiffness results in faster pulse wave propagation through the arteries (i.e., a decrease in arterial PTT) and an increase in arterial BP \cite{ref_5}. Therefore, there exists an inverse relationship between BP and PTT. Numerous PTT-based BP measurement models have been developed based on this theory. For instance, Poon \textit{et al}. \cite{ref_6} proposed an equation that describes the relationship between the square of the inverse of PTT and BP. Building upon this, Ding \textit{et al}. \cite{ding2015continuous} introduced the PPG intensity ratio (PIR), which reflects changes in blood vessel diameter, to further enhance the PTT model. Additionally, Liu \textit{et al}. \cite{liu2018multi} presented an arteriolar PTT-based model that exhibited superior performance compared to traditional PTT models.

Since PTT models with few parameters may be difficult to accurately reflect the complex variability of BP, data-driven approaches based on multi-parameter fusion have been proposed. In these approaches, one method involves extracting BP-related features from biosignals and employing machine learning techniques to map these features to BP values. For example, Haddad \textit{et al}. \cite{ref_7} extracted features from PPG and its derivatives, constructing a BP estimation model using multiple linear regression. Similarly, Yang \textit{et al}. \cite{yang2020estimation} combined PPG features with PTT and developed BP estimation models utilizing linear regression, random forest, artificial neural networks, and recurrent neural networks. Another data-driven approach is end-to-end deep learning, which directly estimates BP from raw biosignal waveforms, eliminating the need for manual feature extraction. For instance, Zabihi \textit{et al}. \cite{zabihi2022bp} proposed a BP estimation model that combines causal dilation convolution and residual concatenation to extract deep features from ECG and PPG signals. Kim \textit{et al}. \cite{ref_8} proposed a U-Net network incorporating self-attention for estimating BP from PPG signals.

\subsection{LLMs for Health Tasks Using Time Series Data}
LLMs for health tasks involving time series data can be examined from two main perspectives: LLMs for time series analysis and LLMs customized for health-related tasks. In terms of LLMs for time series analysis, a crucial aspect is the techniques employed to process the time series data, considering that LLMs are typically trained on text data. Existing works in this area can be broadly categorized into two groups. The first category directly transforms the numerical input and output into prompts. An example of this approach is the notable work PromptCast \cite{xue2023promptcast}. The second category focuses on reprogramming the input time series data into text representations through cross-modality adaptation, making the numeric data more suitable for LLMs' capabilities. The prominent works in this category include Time-LLM \cite{jin2023time} and TEST \cite{sun2023test}. While there are several existing works in this field \cite{xue2023promptcast, jin2023time, sun2023test, cao2023tempo}, most of them \cite{xue2023promptcast, jin2023time, cao2023tempo} concentrate on tasks related to time series forecasting, which can be applied to other domains such as traffic management, climate modeling, and finance analytics. Few studies \cite{sun2023test} have specifically addressed time series classification or regression in health-related tasks.

Regarding the customization of LLMs for health-related tasks, this area is rapidly expanding in research. Although there are several existing works in this field \cite{ref_10, ref_11, ref_12}, most of them focus on tasks related to the text modality, such as answering health questions and generating health reports. Limited research has explored the integration of mobile and wearable biosignals for health-related tasks. Despite LLMs being advanced and cutting-edge techniques in time series analysis and medical tasks, the combination of harnessing LLMs for health tasks using time series data and leveraging their capabilities in the healthcare field has not been thoroughly explored. Notable studies in this context include Health-Learner \cite{ref_13} and Health-LLM \cite{ref_14}. Health-Learner demonstrated that through only few-shot fine-tuning, an LLM can effectively analyze various physiological and behavioral time-series datasets \cite{ref_13}. Similarly, Health-LLM presented a comprehensive evaluation of several state-of-the-art LLMs using prompting and fine-tuning techniques on four public health datasets and ten health-related tasks \cite{ref_14}. However, to the best of our knowledge, existing studies have yet to explore the performance of LLMs in the task of cuffless BP estimation. Therefore, further investigation is needed to assess how LLMs can enhance the accuracy and efficiency of cuffless BP estimation.

\section{Method}
\begin{figure*}[htbp]
\centerline{\includegraphics[width=0.95\textwidth]{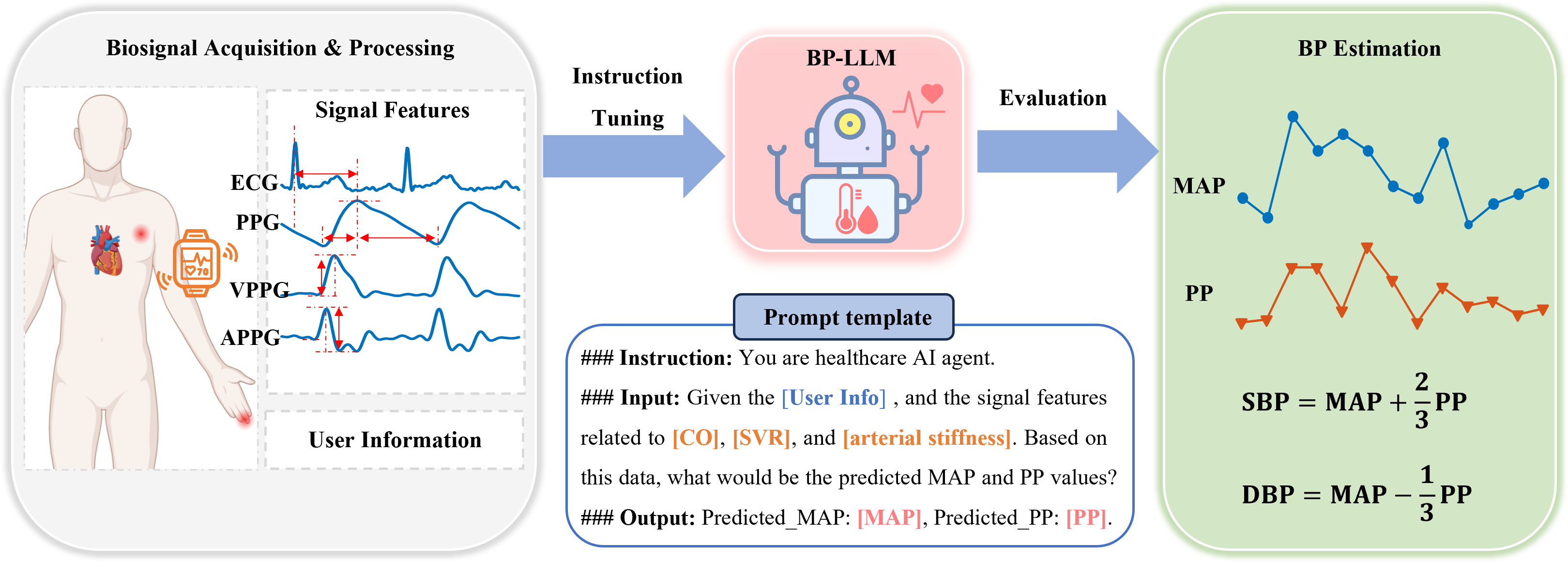}}
\caption{Framework of cuffless BP estimation using LLMs.}
\label{fig2}
\end{figure*}

Previous studies have indicated that the utilization of handcrafted features with physiological meaning for BP estimation can achieve comparable or even better performance than end-to-end deep learning approaches \cite{ref_3, maqsood2022survey}. Building upon this, we extracted relevant features from biosignals and embedded them into textual prompts. These prompts were then used to fine-tune LLMs to adapt them to the BP measurement task. Figure \ref{fig2} illustrates our framework for cuffless BP estimation using LLMs.

\subsection{Signal Feature Extraction}
We used 31 signal features from previous studies \cite{ref_3, ref_15}. These features were extracted from ECG, PPG, the first derivative of PPG (VPPG), and the second derivative of PPG (APPG) signals. Based on their physiological significance, these features can be categorized into those related to cardiac output, peripheral resistance, and arterial stiffness, as listed in Table \ref{tab1}. Detailed definitions and calculations of these features can be found in Appendix A.

\begin{table*}[t]
  \centering
  \caption{Definitions of features.}
  \label{tab1}
  \begin{tabular}{clcccc}
    \toprule
    \multirow{2}{*}{Physiological significance} & \multirow{2}{*}{Features} & \multicolumn{4}{c}{Signal used} \\
    \cmidrule(lr){3-6}
    & & ECG & PPG & VPPG & APPG \\
    \midrule
    \multirow{2}{*}[-1ex]{\parbox{2cm}{\centering Arterial stiffness-related}} 
    & PTT: Time span between ECG R peak to PPG valley point & $\checkmark$ & $\checkmark$ & . & . \\
    & PTT: Time span between ECG R peak to VPPG peak point & $\checkmark$ & . & $\checkmark$  & .\\
    & PTT: Time span between ECG R peak to PPG peak point & $\checkmark$ & $\checkmark$ & . & .  \\
    \midrule
    \multirow{6}{*}[-1ex]{\parbox{2cm}{\centering Cardiac output (CO)-related}} 
    & Ascending time & . & $\checkmark$ & $\checkmark$ & $\checkmark$ \\
    & Descending time & . & $\checkmark$ & $\checkmark$ & $\checkmark$ \\
    & Large artery stiffness index & . & $\checkmark$ & . & . \\
    & Pulse width & . & $\checkmark$ & . & . \\
    & Pulse rate & . & $\checkmark$ & . & . \\
    & Pulse intensity rate & . & $\checkmark$ & . & . \\
    \midrule
    \multirow{6}{*}[-1ex]{\parbox{2cm}{\centering Peripheral resistance (PR)-related}} 
    & Ascending slope & . & $\checkmark$ & $\checkmark$ & $\checkmark$ \\
    & Ascending area & . & $\checkmark$ & $\checkmark$ & $\checkmark$ \\
    & Ascending intensity difference & . & $\checkmark$ & $\checkmark$ & $\checkmark$ \\
    & Descending slope & . & $\checkmark$ & $\checkmark$ & $\checkmark$ \\
    & Descending area & . & $\checkmark$ & $\checkmark$ & $\checkmark$ \\
    & Descending intensity difference & . & $\checkmark$ & $\checkmark$ & $\checkmark$ \\
    \bottomrule
    \multicolumn{6}{l}{\makecell[l]{VPPG and APPG indicate the first and second derivative of PPG, respectively.}}
  \end{tabular}
\end{table*}

\subsection{Prompt Design and Instruction Tuning}

\subsubsection{\textbf{Prompt Design}}
Prompts are critical for pre-training LLMs to adapt to new tasks. To adapt the pre-trained LLMs for the BP estimation tasks, we designed stepwise context-enhanced prompts based on domain knowledge, as shown in Table \ref{tab2}. Specifically: \textbf{1) Basic prompt}: We designed a basic prompt that only uses the signal features. The values of the signal features (numerical data) were embedded in a text format, following previous studies  \cite{ref_13,ref_14}. \textbf{2) With BP domain knowledge}: BP readings typically include four values: systolic blood pressure (SBP), diastolic blood pressure (DBP), mean arterial pressure (MAP), and pulse pressure (PP), with SBP and DBP being the most widely used. MAP and PP can be calculated from SBP and DBP as follows: $MAP=(SBP+2DBP)/3$ and $PP=SBP-DBP$. Physiologically, MAP is primarily influenced by cardiac output and peripheral resistance, whereas PP is mainly affected by arterial stiffness \cite{ref_16}. Based on this domain knowledge, we embedded the signal features into the prompt template in a type-value format, and fine-tuned LLMs to estimate MAP and SBP. Then, the estimates of SBP and DBP were calculated from the estimates of MAP and PP, i.e., $SBP=MAP+2PP/3$ and $DBP=MAP-PP/3$. \textbf{3) With BP domain knowledge and user information}: Previous studies have demonstrated that adding user information can significantly improve the accuracy of BP estimation. Here, we included four user-specific factors: age, gender, height, weight, and a history of hypertension. Appendix B provides an example of combining all contexts in the prompt.

\begin{table*}[h]
  \caption{Stepwise knowledge-enhanced strategy for BP estimation prompts.}
  \label{tab2}
  \renewcommand{\arraystretch}{1.5} 
  \begin{tabular}{>{\centering\arraybackslash}m{0.2\linewidth} >{\raggedright\arraybackslash}m{0.8\linewidth}}
    \toprule
    Context & Prompt \\
    \midrule
    Basic \newline (Signal Features)  &
    The physiological features are {\color{blue}{{[}physiological signal features{]}}}. Based on these data, what would be the predicted MAP and PP values?\\
    \midrule
    With BP domain knowledge \newline (Signal Features + \newline BP Knowledge) &
    Mean arterial pressure (MAP) represents the average blood pressure during a cardiac cycle and is influenced by cardiac output and peripheral resistance. Pulse pressure (PP) is the difference between systolic and diastolic blood pressure and is correlated with arterial stiffness. The physiological features associated with cardiac output are {\color{blue}{{[}cardiac output related features{]}}}, peripheral resistance are {\color{blue}{{[}peripheral resistance related features{]}}}, and arterial stiffness are {\color{blue}{{[}arterial stiffness related features{]}}}. Based on these data, what would be the predicted MAP and PP values?\\
   \midrule
   With BP domain knowledge \newline and user information \newline (Signal Features + \newline BP domain Knowledge + \newline User Information) & 
    Mean arterial pressure (MAP) represents the average blood pressure during a cardiac cycle and is influenced by cardiac output and systemic vascular resistance. Pulse pressure (PP) is the difference between systolic and diastolic blood pressure and is correlated with arterial stiffness. Given the user's profile: age: {\color{orange}{{[}age{]}}} years old, gender: {\color{orange}{{[}gender{]}}}, height: {\color{orange}{{[}height{]}}} cm, weight: {\color{orange}{{[}weight{]}}} kg, history of hypertension: {\color{orange}{{[}yes or no{]}}}. The physiological features associated with cardiac output are {\color{blue}{{[}cardiac output related features{]}}}, peripheral resistance are {\color{blue}{{[}peripheral resistance related features{]}}}, and arterial stiffness are {\color{blue}{{[}arterial stiffness related features{]}}}. Based on these data, what would be the predicted MAP and PP values?\\
  \bottomrule
  \multicolumn{2}{l}{\makecell[l]{{\color{blue}Blue} refers to physiological signal features, {\color{orange}orange} to user information.}}
\end{tabular}
\end{table*}

\subsubsection{\textbf{Instruction Tuning}}
Instruction tuning is a technique that enhances LLMs by incorporating explicit instructions during the fine-tuning process. This approach guides the model's behavior and improves its performance in specific tasks or domains. This paper employs instruction tuning to gain a profound understanding of physiological features, mechanisms, and context, enhancing its capability to generate accurate and contextually appropriate responses.


\subsection{Models}

\subsubsection{\textbf{LLM-based Models}}
In this study, we conducted experiments using 10 open-source LLMs with varying sizes and pre-training goals. The models used were as follows:
\begin{itemize}

\item {\texttt{\textbf{Gemma-7B}} \cite{ref_17}}: A lightweight LLM developed by Google, based on the same technology as the Gemini models. It is well-suited for various text generation tasks, such as question-answering, summarization, and inference. Its compact size allows for deployment on consumer-grade hardware like laptops and desktops.
\item{\texttt{\textbf{Mistral-7B}} \cite{ref_18}}: A pre-trained generative text model with 7 billion parameters, created by the Mistral AI team. It incorporates grouped-query attention to ensure faster inference and utilizes sliding window attention for efficient handling of sequences of any length while minimizing inference cost. Being open source, users can easily download and deploy it in diverse environments.
\item{\texttt{\textbf{Yi-6B}} \cite{ref_9}}: An open-source LLM developed by 01.AI, trained from scratch on an extensive corpus of 3 trillion tokens. It supports both English and Chinese languages and exhibits high performance in several tasks, including Multi-Modal Language Understanding and Common-sense Reasoning. Additionally, it enables scaling up to 32K sequence lengths during inference.
\item{\texttt{\textbf{MedAlpaca-7B}} \cite{ref_10}}: An advanced LLM focused on the biomedical field, based on Stanford Alpaca and AlpacaLoRA. It was trained on an extensive collection of medical texts, enabling efficient handling of medical question-answering and medical dialogue tasks.
\item {\texttt{\textbf{LLaMA2-7B}} \cite{ref_19}}: An open-source LLM with 7 billion parameters launched by Meta in 2023. It has been pre-trained on massive amounts of data and has demonstrated excellent performance in multiple benchmarks, providing a powerful foundation for tasks such as text generation and question-answering.
\item{\texttt{\textbf{LLaMA3-8B\footnote{https://ai.meta.com/blog/meta-llama-3/}}}}: Meta AI's open-source LLM released in April 2024, pre-trained on over 15 trillion tokens collected from publicly available sources. The training dataset is seven times larger than that used for LLaMA2, and it includes four times more code, leading to improved performance on a wide array of AI benchmarks.
\item{\texttt{\textbf{Qwen2-7B}} \cite{ref_20}}: A generic LLM developed by Alibaba Cloud, released in June 2024. Qwen2 has generally surpassed most open-source models and demonstrates remarkable capabilities in understanding complex instructions, language understanding, multilingual capability, coding, reasoning, memorizing, and arithmetic problem-solving.
\item{\texttt{\textbf{PalmyraMed-20B}} \cite{ref_11}}: An LLM that has been further trained on Palmyra-Large, utilizing a specialized and custom-curated medical dataset. It has demonstrated superior performance on medical knowledge benchmarks, including PubMedQA and MedQA. The primary objective of this model is to enhance performance in medical dialogue-related tasks.
\item {\texttt{\textbf{PMCLLaMA-13B}} \cite{ref_21}}: An LLM initialized from LLaMA2-13B and further pre-trained with a medical corpus consisting of 4.8 million biomedical academic papers and 30,000 medical textbooks. It incorporates data-centric knowledge injection through the integration of this corpus. Additionally, it undergoes comprehensive fine-tuning to align with medical domain instructions.
\item{\texttt{\textbf{OpenBioLLM-8B\footnote{https://huggingface.co/aaditya/Llama3-OpenBioLLM-8B}}}}: An advanced open-source language model explicitly designed for the biomedical domain. It was developed by Saama AI Lab and released in May 2024. This model leverages cutting-edge techniques to achieve state-of-the-art performance across various biomedical tasks. The developers continuously innovate, research, and develop AI models for healthcare.

\end{itemize}

\subsubsection{\textbf{Baseline Models}}
For comparison purposes, we also utilized traditional machine learning models, including AdaBoost and Decision Tree Regressor (DTR), which have shown promising results in estimating BP. Similar to LLMs, these models were trained using the extracted signal features and user information as inputs on the same dataset. As this study focused on calibration-based BP estimation (see Section 4.2 for more details), a challenging baseline (zero-baseline) was employed for comparison \cite{ref_1}. The zero-baseline assumes no change in BP from the initial calibration and sets the BP estimates to a constant value corresponding to the basal BP.

\section{Experiment}

\subsection{Dataset}
\begin{figure}[htbp]
\centerline{\includegraphics[width=0.5\textwidth]{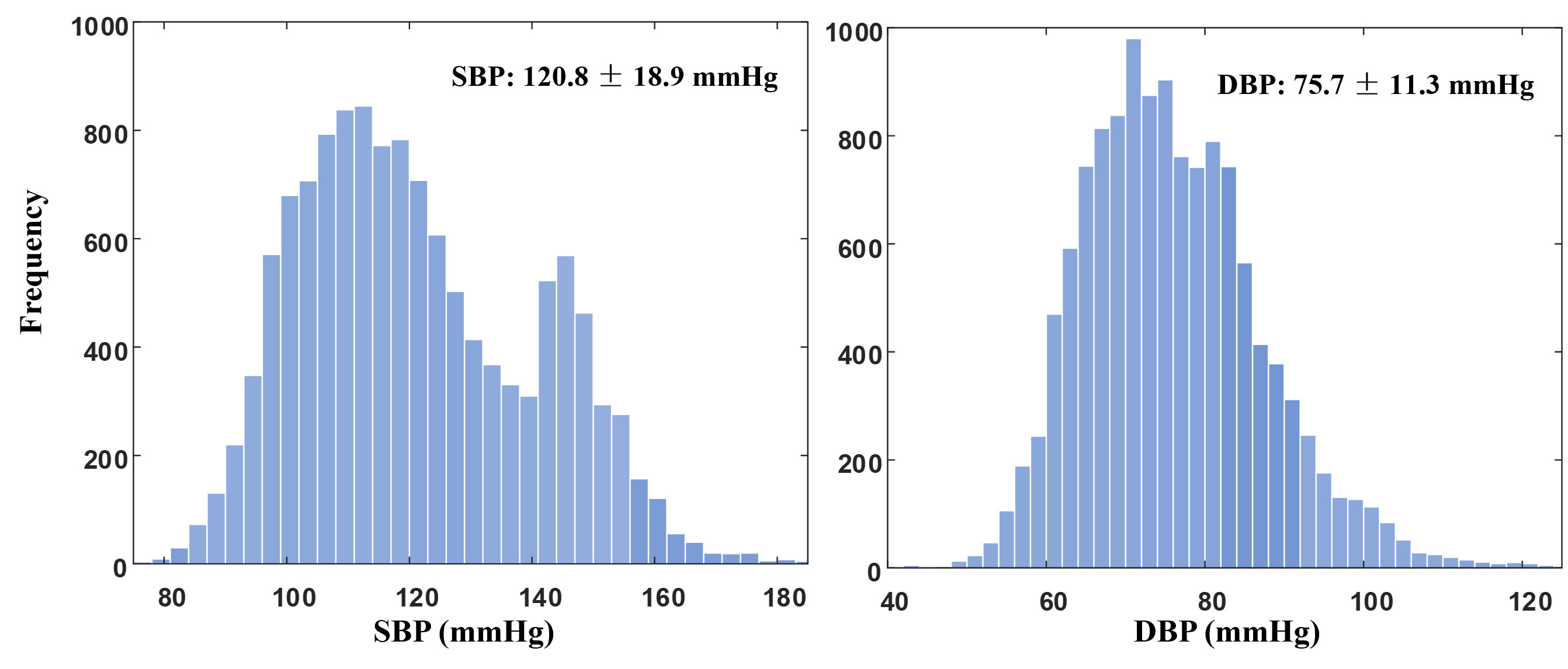}}
\caption{Distribution of SBP and DBP in the CAS-BP dataset.}
\label{fig3}
\end{figure}

We evaluated the LLMs on a large wearable BP dataset: the CAS-BP Dataset\footnote{https://github.com/zdzdliu/CAS-BP}. This dataset, provided by the Shenzhen Institute of Advanced Technology, Chinese Academy of Sciences, contains measurements obtained from 1,272 subjects (mean age 50.16 $\pm$ 12.59 years, 66.82\% female, 25.63\% with hypertension) over 4 days of follow-up (D, D+7, D+14, and D+21). Each measurement includes synchronized ECG and PPG signals with a duration of about 2 minutes, as well as a pair of auscultation SBP and DBP readings. After excluding measurements with poor signal quality, 12,624 measurements were retained for analysis. The sampling rate for both ECG and PPG signals is 1000 Hz. In this dataset, an individual's basal BP was defined as the mean value of the BP measured on day D. Figure \ref{fig3} shows the distribution of SBP and DBP in the CAS-BP dataset. The features listed in Table 1 were extracted on a beat-by-beat basis and then averaged.

\subsection{Implementation Details}
The hardware configuration of our computing system consists of 4 RTX A6000 GPUs with 48 GB of memory each. We experimented with LLMs of various scales, including both general-purpose and domain-specific open-source models. Additional, we performed instruction tuning on these models. The implementations were obtained from Llama Factory \cite{ref_22}.


To evaluate the performance more reliably and robustly, we performed a 5-fold cross-validation for each experiment, with the other configurations remaining unchanged. 

Furthermore, as previous studies have emphasized the necessity of calibrating cuffless BP models, we calibrated both the LLM-based models and baseline models using basal BP, following \cite{ref_3, ref_15}. For each individual, assuming that the model outputs SBP and DBP are denoted as $SBP_{free}$ and $DBP_{free}$, respectively, the calibrated estimates of SBP and DBP, $SBP_{cal}$ and $DBP_{cal}$, can be computed as:

\begin{equation}
    SBP_{cal}=SBP_{free} \cdot \alpha + BaseSBP \cdot (1-\alpha)
\end{equation}
\begin{equation}
    DBP_{cal}=DBP_{free} \cdot \alpha + BaseDBP \cdot (1-\alpha)
\end{equation} where $BaseSBP$ and $BaseDBP$ refer to the individual's basal SBP and basal DBP, respectively. The parameter $\alpha$ was determined experimentally, and in this study, we set it to 0.3 (see Section 5.5 for details).

\section{Results and Discussion}

\subsection{Performance Metrics}
Three metrics specified in major international standards were employed to evaluate the performance of BP estimation models. First, the mean absolute error (MAE) was calculated following the IEEE standard for Wearable Cuffless Blood Pressure Measuring Devices (IEEE 1708) \cite{ref_23}. Second, the mean error (ME) and standard deviation of the error (SDE) were computed in accordance with the American National Standards Institute/ Association for the Advancement of Medical Instrumentation/ International Organization for Standardization (ANSI/AAMI/ISO) standard \cite{ref_24}. The mathematical definitions of these metrics are as follows:
\begin{equation}MAE=\frac{1}{n}\sum_{i=1}^{n}\left|ref_{i}-est_{i}\right|\tag{3}\end{equation}
\begin{equation}ME=\frac{1}{n}\sum_{i=1}^{n}(ref_{i}-est_{i})\tag{4}\end{equation}
\begin{equation}SDE=\sqrt{\frac{1}{n-1}\sum_{i=1}^{n}(ref_{i}-est_{i}-ME)^{2}}\tag{5}\end{equation} where $\left\{est_{1},est_{2},...est_{n}\right\}$ and $\left\{ref_{1},ref_{2},...ref_{n}\right\}$ refer to the estimated and reference BP values, respectively, and $n$ indicates the number of BP measurements.

\subsection{Instruction Tuning Performance}
\begin{figure}[htbp]
\centerline{\includegraphics[width=0.5\textwidth]{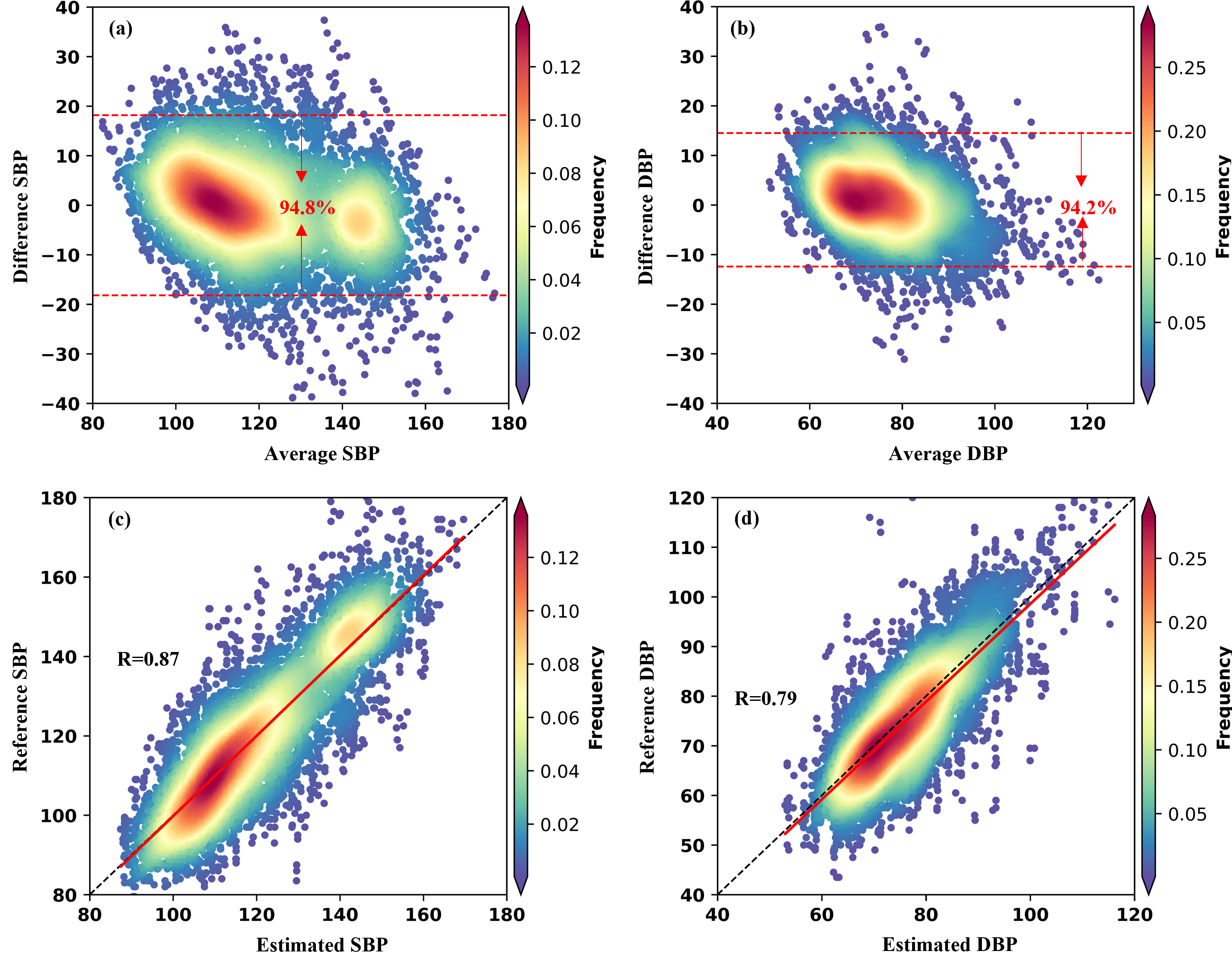}}
\caption{Density Bland–Altman plots of the fine-tuned LLaMA3-8B model for SBP (a) and DBP (b) estimations. Density correlation plots of the fine-tuned LLaMA3-8B model for SBP (c) and DBP (d) estimations.}
\label{fig4}
\end{figure}

Table \ref{tab3} presents the results of the instruction tuning LLMs and baseline methods for BP estimation. As shown, the LLaMA3-8B model achieved the best performance for SBP estimation, while the Qwen2-7B model demonstrated the best performance for DBP estimation. Additionally, the MedAlpaca-7B model, which was obtained by fine-tuning the LLaMA2-7B with medical domain tasks, outperformed the base LLaMA2-7B model. However, the LLaMA3-8B model exhibited superior performance compared to MedAlpaca-7B, likely due to its incorporation of a higher number of parameters. Furthermore, most of the LLMs outperformed the baseline methods. Specifically, when compared to the best-performing baseline (AdaBoost), the LLaMA3-8B model reduced the ME $\pm$ SDE from 1.39 $\pm$ 9.55 mmHg to 0.00 $\pm$ 9.25 mmHg for SBP estimation, and the Qwen2-7B model reduced the ME $\pm$ SDE from 1.62 $\pm$ 7.04 mmHg to 1.29 $\pm$ 6.37 mmHg for DBP estimation. Figure \ref{fig4} shows the density Bland-Altman and correlation plots of the estimated values from the LLaMA3-8B model against the reference values across all folds. The Bland-Altman plots indicate that most of the estimated points for SBP and DBP are within or close to the 95\% limits of agreement, suggesting a close agreement between the predicted BP values and the reference BP values. The correlation plots show that all correlation coefficients are greater than 0.7, indicating a strong correlation between the predicted and reference BP values. These findings suggest that pre-trained LLMs can be effectively adapted to the task of BP estimation through instruction tuning with wearable BP data.

\begin{table*}[t]
  \centering
  \caption{Performance comparison of LLMs and traditional models for BP estimation.}
  \label{tab3}
  \begin{tabular}{llllllll}
    \toprule
    \multirow{3}{*}{Category} & \multirow{3}{*}{Model} & \multicolumn{3}{c}{SBP Estimation (mmHg)} & \multicolumn{3}{c}{DBP Estimation (mmHg)}\\
    \cmidrule(lr){3-5} \cmidrule(lr){6-8}
    &  & MAE (↓) & ME & SDE (↓) & MAE (↓) & ME & SDE (↓)  \\
    \midrule
    \multirow{10}{*}{\makecell{LLM}} & Gemma-7B & 7.32 & -0.01 & 9.47 & 5.86 & 2.76 & 6.99 \\
    & Mistral-7B & 7.15 & -0.50 & 9.29 & 5.31 & 0.99 & 6.89 \\
    & Yi-6b & 7.19 & -0.90 & 9.31 & 5.30 & 0.87 & 6.90  \\
    & MedAlpaca-7B & 7.17 & -0.47 & 9.31 & 5.31 & 1.00 & 6.88 \\
    & LLaMA2-7B & 7.48 & 1.04 & 9.68 & 5.63 & 1.60 & 7.14 \\
    & LLaMA3-8B & \textbf{7.08} & \textbf{0.00} & \textbf{9.25} & 5.31 & 1.07 & 6.86 \\
    & Qwen2-7B &7.30  &1.68  &9.39  &\textbf{5.19}  &\textbf{1.29}  &\textbf{6.37}  \\
    & PalmyraMed-20B &7.51  &1.64  &9.64  &5.40  &1.33  &6.96  \\
    & PMCLLaMA-13B &7.59  &1.75  &9.73  &5.34  &1.36  &6.88  \\
    & OpenBioLLM-8B &7.31  &1.71  &9.43  &5.26  &1.13  &6.87  \\
    \midrule
    \multirow{3}{*}{Baseline} & AdaBoost & 7.42 & 1.39 & 9.55 & 5.51 & 1.62 & 7.04 \\
    & DTR & 7.52 & 1.04 & 9.85 & 5.60 & 1.25 & 7.31 \\
    & Zero-baseline & 7.83 & 1.37 & 10.46 & 5.91 & 1.60 & 7.80 \\
    \bottomrule
  \end{tabular}
\end{table*}

\subsection{Importance of Different Contexts in Prompt Designing for BP Estimation}
We conducted ablation experiments to evaluate the effect of different knowledge enhancement strategies using the LLaMA3-8B model as an example. The results are presented in Table \ref{tab4}. It can be observed that the performance of the model improves with the incorporation of knowledge context. For instance, compared to the basic signal feature, the addition of BP domain knowledge context led to reductions of 1.03\% and 2.0\% in terms of MAE and SDE for SBP estimations, respectively. Furthermore, the inclusion of user information resulted in even greater enhancements, with MAE and SDE reductions of 8.90\% and 9.25\%, respectively. These results demonstrate that incorporating BP domain-specific knowledge and user-relevant factors in the prompt helps the LLM efficiently navigate its large knowledge base and access relevant information, thereby improving its ability to estimate BP accurately.

\begin{table}[t]
  \centering
  \caption{BP estimation results in different contexts.}
  \label{tab4}
  \resizebox{\linewidth}{!}{%
    \begin{tabular}{ccccc}
      \toprule
      \multirow{3}{*}{Context}  & \multicolumn{2}{c}{SBP Estimation} & \multicolumn{2}{c}{DBP Estimation}\\
      \cmidrule(lr){2-3} \cmidrule(lr){4-5}
      & MAE & SDE & MAE & SDE  \\
      & (\% Reduction) & (\% Reduction) & (\% Reduction) & (\% Reduction)  \\
      \midrule
      \multirow{2}{*}{Basic} & 7.77 & 9.98 & 5.37 & 7.04 \\
      & (-) & (-) & (-) & (-) \\
      \multirow{2}{*}{With BP domain knowledge} & 7.69 & 9.78 & 5.37 & 7.01 \\
      & (1.03\%) & (2.0\%) & (0\%) & (0.43\%) \\
      \multirow{2}{*}{\makecell[c]{With BP domain knowledge\\and user information}} & 7.08 & 9.25 & 5.31 & 6.86 \\
      & (8.90\%) & (7.31\%) & (1.12\%) & (2.56\%) \\
      \bottomrule
    \end{tabular}%
  }
\end{table}

\subsection{Importance of Training Size in Instruction Performance}
\begin{figure}[htbp]
\centerline{\includegraphics[width=0.5\textwidth]{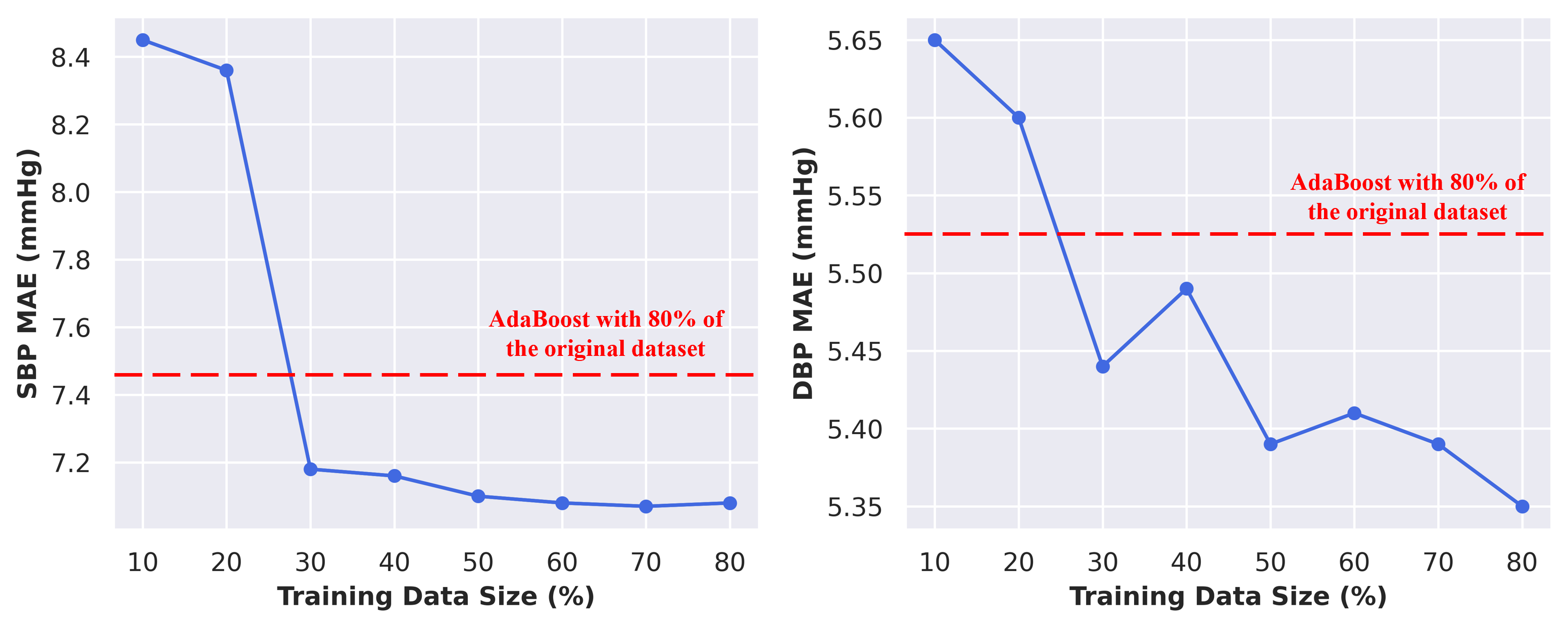}}
\caption{BP estimation of LLaMA-3-8B with different training sizes. The solid lines represent the performance of the fine-tuned LLaMA-3-8B model, while the dashed lines indicate the performance achieved by training AdaBoost with 80\% of the original data, serving as the baseline.}
\label{fig5}
\end{figure}

To understand the required amount of data for instruction tuning, we conducted experiments on the LLaMA3-8B model with instruction tuning varying down-sampled training sizes: 10\%, 20\%, 30\%, 40\%, 50\%, 60\%, 70\%, and 80\% of the original dataset. The overall results are presented in Figure \ref{fig5}. Notably, even with only 30\% of the original dataset, the fine-tuned model already outperforms the baseline model (AdaBoost trained on 80\% of the original data) for both SBP and DBP estimation. This finding provides guidance for instruction tuning when data and computing resources are limited. As expected, we observe an increasing trend in performance with more training data.

\subsection{Effect of the Parameter $\alpha$}
\begin{figure}[htbp]
\centerline{\includegraphics[width=\columnwidth]{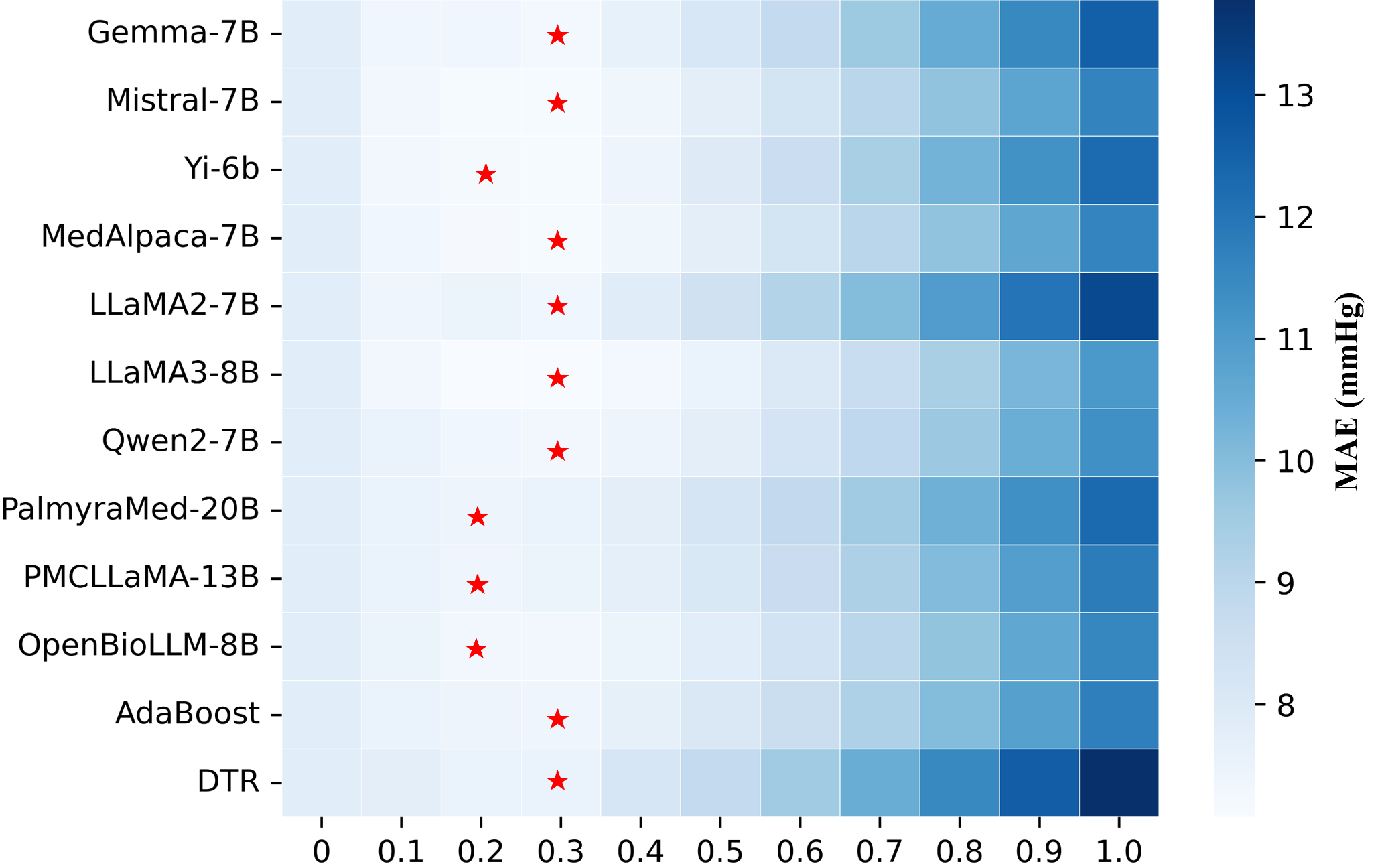}}
\caption{BP estimation models for SBP's MAE at different $\alpha$ values.}
\label{fig6}
\end{figure}

Using the estimation of SBP as an example, we evaluated the performance of the models under different values of the parameter $\alpha$ in Equations (1) and (2). The $\alpha$ value was increased from 0 to 1 in steps of 0.1. The results were presented in Figure \ref{fig6}, with the optimal $\alpha$ value for each model marked by a star symbol. As depicted, most of the models achieved the lowest MAE when $\alpha$=0.3, while a few attained the lowest MAE at $\alpha$=0.2. However, statistical analysis revealed no significant difference in performance between $\alpha$=0.2 and $\alpha$=0.3 across all the evaluated models. Therefore, we determined $\alpha$=0.3 as the optimal value.

\subsection{Limitations and Future Works}
While the results are promising, there are some limitations to consider. The present study evaluated the performance of LLMs on a single wearable dataset and utilized handcrafted features extracted from the signals as inputs to the LLMs, without assessing their performance in BP estimation from raw signal waveforms. To address these limitations, future research efforts will include 1) evaluating the performance of LLMs on multiple wearable datasets, 2) exploring the potential of LLMs to directly estimate BP from raw signal waveforms, and 3) investigating the decision-making process of LLMs to enhance the interpretability for BP estimation.

\section{Conclusion}

This paper proposes a novel application of using LLMs for cuffless BP estimation from wearable biosignals. We extracted rich physiological features from wearable ECG and PPG signals and integrated them with BP domain-specific knowledge and user information to design context-enhanced prompt templates. Subsequently, we fine-tuned the LLMs with these prompts to adapt them for the BP estimation task. The results obtained from instruction tuning ten LLMs on a wearable dataset comprising 1,272 subjects demonstrate that the performance of the LLMs surpasses that of baseline models. Additionally, our ablation studies further emphasize the importance of incorporating BP domain knowledge and user information into the prompts. These findings demonstrate the feasibility and potential of LLMs in enhancing the performance of cuffless BP estimation.


\bibliographystyle{unsrt}
\bibliography{references.bib}

\appendix

\section{Appendix A. Calculations of Signal Features}

\begin{figure}[htbp]
\centerline{\includegraphics[width=0.5\textwidth]{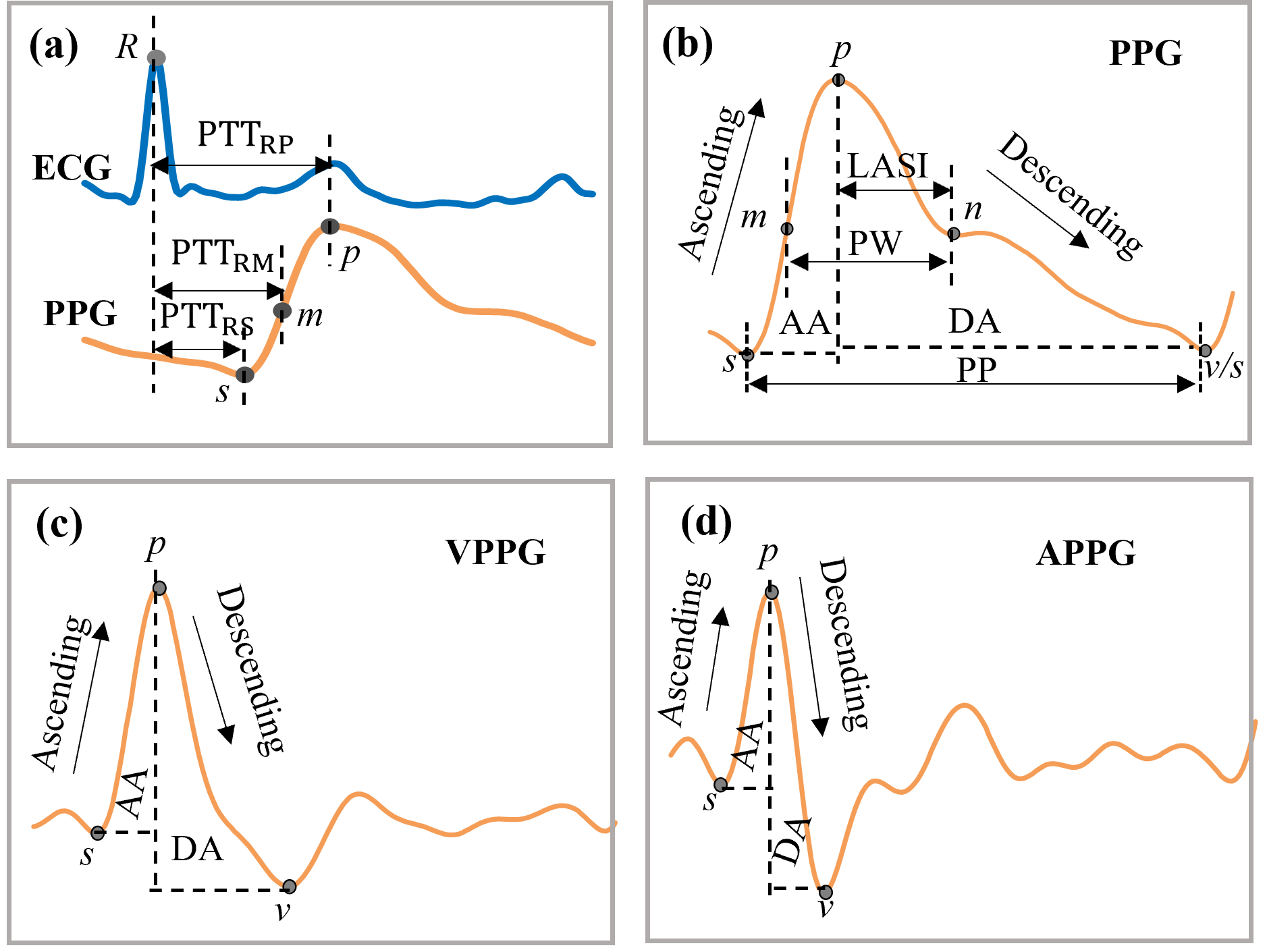}}
\caption{(a) Detection of relevant feature points in ECG and PPG for PTT calculation. (b) Detection of relevant feature points in PPG for PPG-based features calculation. (c) Detection of relevant feature points in VPPG for VPPG-based features calculation. (d) Detection of relevant feature points in APPG for APPG-based features calculation.}
\label{fig7}
\end{figure}

The raw ECG and PPG signals were band-pass filtered from 0-40 Hz and 0-20 Hz, respectively. Subsequently, we performed beat-by-beat feature point detection on the filtered signals to calculate beat-by-beat signal features. The feature points included the R-peak ($R$) of ECG, the offset ($s$), maximum slope ($m$) on the upward rise, peak ($p$), dicrotic notch ($n$), and valley ($v$) peaks of PPG, as well as the offset ($s$), peak ($p$), and valley ($v$) peaks of the first derivative of the PPG (VPPG), and the offset ($s$), peak ($p$), and valley ($v$) peaks of the second derivative of the PPG (APPG), as shown in Figure \ref{fig7}. The calculation of each feature is detailed in Table \ref{tab5}.

\begin{table*}[h]
  \caption{Calculation of signal features.}
  \label{tab5}
  \begin{tabular}{p{0.2\linewidth}p{0.4\linewidth}p{0.4\linewidth}}
    \toprule
    \multicolumn{1}{c}{\centering \textbf{Number}} & \multicolumn{1}{c}{\centering \textbf{Feature}} & \multicolumn{1}{c}{\centering \textbf{Calculation}} \\
    \midrule
    \centering 1  & $\rm PTT_{RV}$ & Time span between ECG $R$ peak and PPG $s$ point\\
    \centering 2  & $\rm PTT_{RM}$ & Time span between ECG $R$ peak and PPG $m$ point\\
    \centering 3  & $\rm PTT_{RP}$ & Time span between ECG $R$ peak and PPG $p$ point\\
    \centering 4-6  & Ascending time of PPG, VPPG and APPG & Time span between points $s$ and $p$\\
    \centering 7-9  & Ascending slope of PPG, VPPG and APPG & Slope between points $s$ and $p$\\
    \centering 10-12  & Ascending area of PPG, VPPG and APPG & Area below the curve surrounded by points $s$ and $p$\\
    \centering 13-15  & Ascending intensity difference of PPG, VPPG and APPG & Amplitude difference between points $p$ and $s$\\
    \centering 16-18  & Descending time of PPG, VPPG and APPG & Time span between points $p$ and $v$\\
    \centering 19-21  & Descending slope of PPG, VPPG and APPG & Slope between points $p$ and $v$\\
    \centering 22-24  & Descending area of PPG, VPPG and APPG & Area below the curve surrounded by points $p$ and $v$\\
    \centering 25-27  & Descending intensity difference of PPG, VPPG and APPG & Amplitude difference between points $p$ and $v$\\
    \centering 28  & Large artery stiffness index of PPG & Time span between points $p$ and $n$\\
    \centering 29  & Pulse width of PPG & Time span between points $m$ and $n$\\
    \centering 30  & Pulse rate of PPG & Time span between points $s$ and $v$\\
    \centering 31  & Pulse intensity rate of PPG & Ratio of $p$ point intensity to $s$ point intensity\\
  \bottomrule
\end{tabular}
\end{table*}

\section{Appendix B. Example of prompts for Instruction Tuning LLMs}
An example of prompts for instruction tuning LLMs is shown in Table \ref{tab6} below.
\begin{table*}[h]
  \caption{An example of prompts for instruction tuning LLMs.}
  \label{tab6}
  \begin{tabular}{p{\linewidth}}
    \toprule
    \textbf{Instruction:} You are a personalized healthcare agent trained to predict mean arterial pressure and pulse pressure based on user information and physiological features calculated from electrocardiogram and photoplethysmogram signals.\\
    \midrule
    \textbf{Input:} Mean arterial pressure (MAP) represents the average blood pressure during a cardiac cycle and is influenced by cardiac output and peripheral resistance. Pulse pressure (PP) is the difference between systolic and diastolic blood pressure and is correlated with arterial stiffness. Given the user's profile: age: 56 years old, gender: female, height: 155.0 cm, weight: 54.0 kg, history of hypertension: no. The physiological features associated with cardiac output are [0.16, 0.51, 0.18, 0.29, 0.66, 0.1, 0.18, 0.07, 0.08], peripheral resistance are [2.64, 16.83, 231.92, 2.64, 5.2, 705.58, 0.89, 0.54, 0.04, 0.39, 1.77, 0.04, 0.23, 2.58, 0.01, 0.03, 0.08, 0.01, 0.05], and arterial stiffness are [0.18, 0.24, 0.33]. Based on these data, what would be the predicted MAP and PP values?\\
    \midrule
    \textbf{Response:} Predicted\_MAP: 86.0 mmHg, Predicted\_PP: 36.0 mmHg.\\
    \bottomrule
\end{tabular}
\end{table*}

\end{document}